\documentclass[prb,aps,preprint,showpacs,superscriptaddress]{revtex4-2}
\usepackage{graphicx}  
\usepackage{dcolumn}   
\usepackage{bm}        
\usepackage{amssymb}
\usepackage{mathtools}
\usepackage{adjustbox}
\usepackage{lipsum}
\usepackage[english]{babel}
\usepackage{hyperref}
\hypersetup{
	colorlinks=true,
	linkcolor=blue,
	filecolor=blue,      
	urlcolor=blue,
}
\begin{document}
\title{High pressure ferroelectric-like semi-metallic state in $Eu-$doped $BaTiO_3$ }
\author{Mrinmay Sahu}
\affiliation {National Centre for High Pressure Studies, Department of Physical Sciences, Indian Institute of Science Education and Research Kolkata, Mohanpur Campus, Mohanpur – 741246, Nadia, West Bengal, India}
\author{Bishnupada Ghosh}
\affiliation {National Centre for High Pressure Studies, Department of Physical Sciences, Indian Institute of Science Education and Research Kolkata, Mohanpur Campus, Mohanpur – 741246, Nadia, West Bengal, India}
\author{Boby Joseph}
\affiliation {Elettra-Sincrotrone Trieste, S. S. 14 km 163.5, 34149 Basovizza, Trieste, Italy}
\author{Goutam Dev Mukherjee}
\email [Corresponding Author:]{ goutamdev@iiserkol.ac.in}
\affiliation {National Centre for High Pressure Studies, Department of Physical Sciences, Indian Institute of Science Education and Research Kolkata, Mohanpur Campus, Mohanpur – 741246, Nadia, West Bengal, India}
\date{\today}
\begin{abstract}
We have conducted a detailed high-pressure (HP) investigation on $Eu-$doped $BaTiO_3$ using angle-resolved x-ray diffraction, Raman spectroscopy, dielectric permittivity and dc resistance measurements. The x-ray diffraction data analysis shows a pressure-induced structural phase transition from the ambient tetragonal to the mixed cubic and tetragonal phase above 1.4 GPa. The tetragonality of the sample due to the internal deformation of the $TiO_6$ octahedra caused by charge difference from Eu doping cannot be lifted upon by pressure. Softening, weakening, and disappearance of low-frequency Raman modes indicate ferroelectric tetragonal to the paraelectric cubic phase transition. But the pressure-induced increase in the intensity of [E(LO), A1(LO)] and the octahedral breathing modes indicate the local structural inhomogeneity remains in the crystal and is responsible for spontaneous polarization in the sample. Low-frequency electronic scattering response suggests the pressure-induced carrier delocalization, leading to a semi-metallic state in the system. Our HP dielectric constant and dc resistance data can be explained by the presence of pressure-induced localized clusters of microscopic ferroelectric ordering. Our results suggest HP phase coexistence leads to a ferroelectric-like semi-metallic state in $Eu-$doped $BaTiO_3$ under the extreme quantum limit.             
\end{abstract}
\maketitle
\section{Introduction}
 Perovskite $BaTiO_3$ (BT) is one of the most extensively investigated lead-free ferroelectric materials since its discovery more than 70 years ago \cite{acosta2017batio3}. It has potential applications in transducers, multi-layer ceramic capacitors, piezoelectric devices, optoelectronic elements, high temperature sensors and random access memories etc. \cite{zhang2006ferroelectric, ismail2016review, zulueta2017defect, Jinghui Gao, A.S.Bhalla}. At ambient pressure, with decreasing temperature, BT exhibits a series of phase transitions: paraelectric cubic $\xrightarrow{\sim 396K}$ ferroelectric tetragonal  $\xrightarrow{\sim 278K}$ orthorhombic  $\xrightarrow{\sim 183K}$ rhombohedral phase \cite{Y.Iwazaki}. The Jahn-Teller distortion in $Ti-O_6$ octahedra is responsible for the macroscopic spontaneous electrical polarization in $BaTiO_3$ \cite{S.Lenjer, S.Mukherjee}. The disappearance of ferroelectric polarization in $BaTiO_3$ is of great importance for piezoelectric device applications and for understanding the origin of ferroelectric ordering in the system \cite{acosta2017batio3}. Pressure-dependent different phase transitions associated with the transformation of ferroelectric to para-electric ordering have been observed at about 2, 5 and 9 GPa, and remain controversial till now \cite{venkateswaran1998high, H.Zhang, M.Malinowski, P.Pruzan, A.K.Sood, J.P.Iti, L.Ehm}. Ferroelectricity in $BaTiO_3$ can be easily removed by applying pressure, introducing electron doping, impurities and defects or by the surface effects in fine particles and thin films. \cite{poojitha2019effect, A.S.Bhalla}. $BaTiO_3$ is a band insulator with a band gap of 3.2 eV \cite{zhang2016photoluminescence, J.Li}. It is observed that ferroelectric $BaTiO_3$ shows a transition from insulating to the metallic, semiconducting or semi-metallic state after electron doping or by creating oxygen vacancies \cite{K.Page, T.Kolodiazhnyi}. The groups, Fritsch et al. \cite{V.Fritsch} and Marucco et al. \cite{J.F.Marucco} show that metallic conductivity appears in $BaTiO_3$ due to the incorporation of $La^{3+}$ and $Nb^{5+}$ dopants at $Ba^{2+}$ and $Ti^{4+}$ ion positions, respectively. The pressure-induced insulator-to-metal transition has also been observed in hexagonal $BaTiO_{3-\delta}$ \cite{Y.Akishige}. One would expect pressure can induce the insulator-to-metal or semi-metal transition in electron-doped $BaTiO_3$, and the ferroelectric phase would disappear under high pressure because of the screening of long-range electrostatic interactions, responsible for the appearance of ferroelectricity \cite{Y.Iwazaki} or the presence of a quantum paraelectric phase due to quantum fluctuation that prevents the formation of dipole \cite{muller1979srti}. However, the ferroelectric-like metallic state is also observed in electron-doped $BaTiO_3$  due to the coexistence of screening of long-range electrostatic interactions by the charge carriers and the local effects \cite{K.Page,J.Fujioka}. High-pressure studies on this electron-doped $BaTiO_3$ are not reported to the best of our knowledge.\\
 Recently the rare earth (RE) doped $ABO_3$ type perovskites have received attention because of their ability to modify the electrical conductivity, optical properties, and thermoelectric properties and to introduce the magnetic ordering \cite{zulueta2017defect, T.Kolodiazhnyi, C. Jiang}. The incorporation of RE ions in $BaTiO_3$ controls the curie temperature, grain growth, and electrical resistance; lowers the dissipation factor, and enhances the dielectric response \cite{ismail2016review, J.Li, V.Fritsch}. However, as mentioned previously, the metallic conductivity can also be achieved in $RE^{3+}$ doped $BaTiO_3$, and it can increase under high pressure. The induced magnetic ordering due to the incorporation of rare earth ions can be modified under pressure. One may also expect the ferroelectric-like metallic or semi-metallic state in $RE^{3+}$ doped $BaTiO_3$ at high pressure. To address these issues, it is essential to understand the structural and electrical properties of $RE^{3+}$ doped $BaTiO_3$ at high pressure, which is not yet studied. In this work, we have synthesized rare earth $Eu^{3+}$ doped $BaTiO_3$ (EBTO) and investigated the pressure-dependent structural and electronic properties by means of Raman spectroscopy, x-ray diffraction, dielectric constant, and dc resistance measurements. Our pressure-dependent x-ray diffraction data show a structural transition from ambient tetragonal-P4mm to the mixed phase tetragonal- P4mm ($25\%$) + cubic- Pm$\bar{3}$m ($75\%$) above 1.4 GPa. We have observed $TiO_6$ octahedral deformation with pressure in the tetragonal phase. Our high-pressure Raman data supports the phase coexistence above 2 GPa. A structural reorientation has also been observed near 10 GPa. Furthermore, the dc resistance and the dielectric measurements reveal that the high-pressure paraelectric states are fragile against the presence of localized microscopic ferroelectric ordering in EBTO.            
\section{Experimental Details}
\subsection{Material synthesis and structural characterizations}
The $Ba_{1-x}Eu_{x}TiO_3$ ($x=0.05$) powder sample was synthesized through the conventional solid state reaction route \cite{kumari2021systematic,zhang2016photoluminescence,de2020large}. The starting reagents $BaCO_3$ (99.5\% purity), $TiO_2$ (99.5\% purity) and $Eu_2O3$ (99.9\% purity) procured from Sigma-Aldrich were weighed in appropriate stoichiometric ratios and appropriately mixed in an agate mortar. The mixed fine powder was calcined at 1200 $^{\circ}$C for 12 hours. After proper grinding the resulting calcined powder was pressed into a pellet form. The pellet was sintered in air at 1400 $^{\circ}$C for 12h to get the final product. Our ambient synchrotron X-ray powder diffraction (XRPD) data confirmed the single-phase tetragonal perovskite crystal structure having tetragonality (c/a) 1.0067(6) (discussed in the next section).
\subsection{High-pressure x-ray powder diffraction measurements}
High-pressure XRPD measurements of polycrystalline EBTO at room temperature were carried out at the XPRESS beamline in ELETTRA synchrotron source in Trieste, Italy. For sample loading inside the DAC, we followed the procedure described by Joseph et al. and Lotti et al. \cite{joseph2018coexistence, lotti2020single}. The incident high energy x-ray radiation of wavelength 0.4957 $\AA$ was collimated to about 25 $\mu$m. The distance between the sample and the detector was calibrated using the x-ray diffraction patterns of the standard sample $CeO_2$. The diffraction data were recorded using a PILATUS 3S 6M detector. DIOPTAS software was used to integrate all the diffraction patterns to $2\theta$ vs intensity profiles \cite{prescher2015dioptas} and then analyzed using CRYSFIRE. GSAS software was used to perform the Lebail, and Rietveld refinements \cite{toby2001expgui}.  
\subsection{High-pressure Raman measurements}
We have carried out pressure-dependent Raman spectroscopy measurements using a piston-cylinder type diamond anvil cell (DAC) from Easy Lab Co. (UK). The diamond culet size was 300 $\mu$m. A pre-indented (indented thickness $\sim$ 45 $\mu$m) steel gasket with a central hole diameter 100 $\mu$m was placed accurately in between two oppositely faced diamond culets to make a proper sample chamber. Fine EBTO powder was loaded inside the sample chamber. 4:1 methanol-ethanol mixture was used to achieve the hydrostatic condition inside the pressure cell. A few ruby chips of the approximate size, 3-5 $\mu$m, were placed along with the powder sample for pressure measurements using ruby fluorescence technique \cite{mao1986calibration,chijioke2005ruby}.
High-pressure Raman spectroscopy measurements were conducted using a confocal micro-Raman system (Monovista from S\&I GmbH) in back scattering geometry. The 532 nm radiation of the Cobalt-samba diode pump laser was used as an excitation source. An infinitely corrected 20X objective lens with a large working distance and laser spot diameter of about 5 $\mu$m was used to collect the Raman signal. A grating of 1500 grooves/mm was used to disperse the scattered light with a spectral resolution of about 2 $cm^{-1}$. For a detailed analysis of the low-frequency electronic contribution near the elastic line, we have recorded Raman spectra using both the edge and Bragg filters with cut-off near 60 $cm^{-1}$ and 4 $cm^{-1}$, respectively. 
\subsection{Pressure dependent dielectric permittivity, dielectric loss, and dc resistance measurements}
High-pressure dielectric permittivity ($\epsilon (P)$), dielectric loss ($\delta (P)$), and dc resistance ($R_{dc}$) measurements were carried out up to 7.1 GPa using a toroid-anvil (TA) apparatus in a large-volume pressure cell \cite{R. Jana}. We used a 300-ton hydraulic press to achieve a maximum pressure of up to 9 GPa inside the sample chamber. A 3 mm diameter and 1 mm thick pellet of the powder sample was sandwiched in between two copper plates of thickness about 0.1 mm and placed properly inside a Teflon cup. The Teflon cup filled with the sample was placed inside a 5 mm diameter hole drilled in the center of the toroid-shaped pyrophyllite gasket. Two thin steel wires with a diameter of 20 $\mu$m were connected to the copper plates and extracted through the tiny holes in the Teflon cup and, finally, through the side holes of the pyrophyllite gasket. The sample assembly with the gasket was well compressed and locked between the two opposed toroid-shaped anvils for our measurements. We kept the sample assembly for around 12 hours at an initial pressure of 0.5 GPa to ensure good contact between the electrodes and the sample. No deformation of the Teflon cup was observed and ensured the reproducible data in the pressure range of 0-7.1 GPa. Pressure-dependent relative permittivity $\epsilon_r$ of the material was calculated from the measured capacitance using the relation,
\begin{equation}
 	C = \epsilon_0\epsilon_r(A/t)
\end{equation}                                
Where C is the capacitance, A and t are the area of the copper plates and the thickness of the sample, respectively. $\epsilon_0$ is the permittivity in free space. 
\section{Results and Discussion}
\subsection{High-pressure X-ray powder diffraction}
We have performed pressure-dependent x-ray powder diffraction of EBTO up to 15.9 GPa, as shown in Fig. 1. At ambient conditions, all the reflection lines are successfully indexed with the tetragonal space group $P4mm$. The Rietveld analysis of the ambient XRPD pattern, as shown in Fig. 2(a), confirmed the stabilization of the tetragonal phase. The details of the refined structural parameters and the goodness of fit are given in Table I. Our data matches extremely well with the reported values \cite{de2020large}. We have not seen any mixed phase at ambient pressure. The calculated cell parameters from the the best fit are a = 3.9966(5)$\AA$, c = 4.0236(8)$\AA$ and corresponding unit cell volume ($V_0$) is 64.2710(9)$\AA^3$. $V_0$ of undoped $BaTiO_3$ is 64.41$\AA^3$ as reported by Lu et al.\cite{lu2011self}. The decrease in unit cell volume indicates a partial replacement of $Ba^{2+}$ by $Eu^{3+}$ ions. Eu ions can enter both the Ba- and Ti-sites and may form the self-compensation clusters of $Eu_{Ba}^{3+}$ - $Eu_{Ti}^{3+}$ in the $BaTiO_3$ lattice with an expansion of the unit cell volume compared to pure $BaTiO_3$ \cite{lu2011self}. This self-compensation mechanism for the charge balance is excluded due to the reduction of unit cell volume and confirms that the electron doped $Eu:BaTiO_3$ is formed by substituting $Eu^{3+}$ ions at the $Ba^{2+}$ site.\\
\begin{table}[h]
\begin{center}
		\small
	\begin{tabular}{m{5em} m{5em} m{5em} m{5em} m{5em}}
		\hline\hline
		&\multicolumn{4}{c}{Tetragonal (P4mm) phase}\\
		\hline
		Atoms& Site& $x/a$& $y/b$& $z/c$\\ [2ex] 
		\hline
		Ba/Eu& 1a& 0.0000& 0.0000& 0.0000 \\ 
		Ti& 1b& 0.5000& 0.5000& 0.514(2) \\
		O1& 1b& 0.5000& 0.5000& 0.025(4) \\
		O2& 2c& 0.5000& 0.0000& 0.503(2) \\ [3ex]
		&a=3.9966(5)$\AA$,&&c=4.0236(8)$\AA$\\
		&$V_0$=64.2710(9)$\AA^3$\\
		&$R_{wp}$=1.8\% &$R_{p}$=1.4\%\\
		\hline\hline
	\end{tabular}
\caption{\textbf {Reitveld refined structural parameters of the ambient XRPD pattern of EBTO.}}
\end{center}
\end{table}
At about 1.4 GPa and above, all the doublet Bragg peak profiles turn into singlets, as shown in the insets in Fig. 1, suggesting a cubic-perovskite-like structure \cite{garg2013lead}. The XRPD pattern of EBTO at 1.4 GPa could not be matched to the tetragonal (P4mm) phase alone. The cubic (Pm$\bar{3}$m) phase model \cite{rao2013local} also failed to fit the XRPD patterns accurately after 1.4 GPa. Next, we consider a two-phase (tetragonal(P4mm) + Cubic(Pm$\bar{3}$m)) coexistence model, which results in a satisfactory fit with $75\%$ cubic (Pm$\bar{3}$m) and $25\%$ tetragonal (P4mm) phases at 2 GPa. Fig. 2(c) shows that after 2 GPa pressure, both the phase percentages remain nearly constant up to the highest pressure value. To understand the phase coexistence model, we performed a comparative fitting of the XRPD pattern at 2 GPa. We consider three models (i) pure tetragonal(P4mm) (ii) pure cubic (Pm$\bar{3}$m) and (iii) tetragonal (p4mm) + cubic (Pm$\bar{3}$m).The best possible fit patterns with the above-mentioned three models are shown in Fig. 3. Comparatively, the cubic (Pm$\bar{3}$m) model gives a better fit ($\chi^2=1.6, R_p=2.6\%$ and $R_{wp}=4.4\%$) than the pure tetragonal (P4mm) model fit ($\chi^2=2.7, R_p=3.4\%$ and $R_{wp}=5.9\%$). The single phase cubic (Pm$\bar{3}$m) model properly fits the lower angle Bragg profiles but does not match the position and intensity of the higher angle reflection lines accurately. One can notice there are some regions in the diffraction pattern (as highlighted in Fig. 3) where remarkable discrepancies occur between the observed and calculated diffraction patterns after the best possible fit with the tetragonal (P4mm) as well as cubic (Pm$\bar{3}$m) model. For example, the calculated intensity profiles of the observed pseudocubic $\{211\}_c$, $\{300\}_c$, $\{310\}_c$ and $\{311\}_c$ Bragg reflection lines do not match with both the tetragonal (P4mm) and cubic(Pm$\bar{3}$m) model fit, as shown in Fig. 3. The best fit, which takes into account all the diffraction peak intensities and positions most accurately, is found to be tetragonal (P4mm) + cubic (Pm$\bar{3}$m) phase coexistence model. We have performed the Rietveld refinement of the XRPD pattern at 2 GPa with the phase coexistence model, as shown in Fig. 2(b). All the refined parameters at 2 GPa are given in Table II. The volume of the tetragonal unit cell (V=62.9965(3) $\AA^3$) and the cubic unit cell (V=62.9850(4) $\AA^3$) are very close at the phase boundary. The values of the lattice parameters of both phases are also very close, as listed in Table II. The pressure evolution of the lattice parameters for both phases is shown in Fig. 4(a). All lattice parameters show similar compression behavior with pressure. However, the c-axis, in the case of the tetragonal phase, shows slope change at transition pressure and a slight saturation is observed around 9 GPa. In both the phases, the unit cell volume vs pressure (V-P) data obtained after refinement of XRPD patterns at all pressure points are fitted to the third-ordered Birch- Murnaghan (BM) equation of state (EoS) \cite{birch1947finite,angel2014eosfit7c}, as shown in Fig. 4(b).
\begin{equation}
	P(V)=\frac{3 B_{0}}{2}\left[\left(\frac{V_{0}}{V}\right)^{7 / 3}-\left(\frac{V_{0}}{V}\right)^{5 / 3}\right] \times\left\{1+\frac{3}{4}\left(B^{\prime}-4\right)\left[\left(\frac{V_{0}}{V}\right)^{2 / 3}-1\right]\right\}
\end{equation}
$B_0$ and $B'$ are the bulk modulus and its first-order pressure derivative, respectively. $V_0$ is the unit cell volume at ambient conditions. The obtained EOS parameters are: for the tetragonal phase, the bulk modulus $B_0 = 116(3)$ GPa; and for cubic phase $B_0 = 122(4)$ GPa. Our results show a slightly smaller compressibility in the cubic phase.  
\begin{table}
	\begin{center}
		\small
			\begin{tabular}{p{1.5cm} p{1.5cm} p{2cm} p{1.5cm} p{2.5cm} p{1.5cm} p{1.5cm} p{1.2cm} p{1.2cm} p{1.2cm}}
			\hline\hline
			&\multicolumn{3}{c} {Tetragonal (P4mm) phase} &\multicolumn{5}{c}{Cubic (Pm$\bar{3}$m) phase}\\
			\hline
			Atoms& Site& $x/a$& $y/b$& $z/c$&  Site& $x/a$& $y/b$& $z/c$\\[2ex]
			\hline  
			Ba/Eu& 1a& 0.0000& 0.0000& 0.0000 &1a &0.0000 &0.0000 &0.0000 \\ 
			Ti& 1b& 0.5000& 0.5000& 0.5487(3) &1b &0.5000 &0.5000 &0.5000 \\
			O1& 1b& 0.5000& 0.5000& 0.0250(6) &3c &0.5000 &0.5000 &0.0000 \\
			O2& 2c& 0.5000& 0.0000& 0.5031(2)& \\[2ex] 
			&a=3.97685(9)$\AA$,c=3.98325(2)$\AA$ &&&& a=3.978742(2),V=62.9850(4)$\AA^3$\\
			&V=62.9965(3)$\AA^3$,Phase\%=25 &&&& Phase\%=75 \\
			\hline\hline
		\end{tabular}
		\caption{\textbf {Reitveld refined structural parameters of the x-ray diffraction pattern of EBTO at 2 GPa with a two-phase model [trtragonal (P4mm)+ cubic (Pm$\bar{3}$m)].}}
	\end{center}
\end{table}
In the tetragonal phase, the most noticeable change in crystal structure upon applying pressure is the change in all six Ti-O bond lengths in the $TiO_6$ octahedra, as shown in Fig. 5(a). Four equivalent in-plane (\emph{xy-plane}) Ti-O bond lengths decrease linearly (at a rate -0.003 $\AA$/GPa) with pressure up to the highest pressure value. In contrast, the other two Ti-O bond lengths along the c-axis behave in the opposite manner with pressure. The large Ti-O bond distance along the c-axis decreases with pressure. Significant slope changes -0.03 $\AA$/GPa to -0.015 $\AA$/GPa and -0.015 $\AA$/GPa to -0.003 $\AA$/GPa are observed at around 3.1 GPa and 8.5 GPa respectively. On the other hand, the small Ti-O bond distance along the c-axis increase with pressure along with significant slope changes 0.019 $\AA$/GPa to 0.009 $\AA$/GPa and 0.009 $\AA$/GPa to -0.0005 $\AA$/GPa at around 3.1 GPa and 8.2 GPa respectively. The experiments were carried out using methanol-ethanol pressure transmitting medium and hence the pressure inside the DAC can be assumed hydrostatic at these pressure values. Therefore, the effect of non-hydrostatic compression on these anomalies can be ruled out. These dissimilar bond distance variations motivate us to observe the variation of Jahn-Teller distortion under high pressure. The octahedral distortion for the $TiO_6 $ octahedra is defined as,
\begin{equation}
D_{JT} = 1/n\sum_{i=1}^{n}{|l_i-l_{av}|}
\end{equation}
where $l_i$ is the individual bond length and $l_{av}$ is the average Ti-O bond length \cite{stekiel2022pressure}. The octahedral distortion of the $TiO_6$ octahedra reduces from 0.015 $\AA$ at 0.5 GPa to 0.002 $\AA$ at 2 GPa, as shown in Fig. 5(b). The decrease in octahedral distortion is due to the pressure-driven decrease in the tetragonality of the sample and is also corroborated by a decrease in the $c/a$ ratio up to the same pressure. The reduction of the octahedral distortion and hence Jahn-Teller distortion indicate the charge delocalization in the system \cite{huang2017jahn,yoon1998raman}.
Interestingly, above 2 GPa, the octahedral distortion increases with pressure and can be related to phase coexistence (cubic + tetragonal) up to the highest pressure value. Our XRPD analysis shows that 75\% of the tetragonal phase transforms to the cubic structure above 1.4 GPa due to the reduction of octahedral distortion and formation of regular $TiO_6$ octahedra under high pressure. Near about 2 GPa pressure, 25\% unit cell of EBTO maintains its tetragonality due to the internal deformation of the $TiO_6$ octahedra. After 2 GPa, the tetragonal phase percentage remains almost constant up to the highest pressure value, as indicated in Fig. 2(c). The distortion of the $TiO_6$ octahedra increases significantly above 2 GPa for the remaining tetragonal phase, with a significant slope change at about 8.2 GPa and stabilizing to a high value. Therefore, it may be noted that the octahedral deformation produced by the charge difference cannot be lifted by pressure; instead, this deformation increases with pressure. The tetragonality of the system decreases from 1.0042 at 0.5 GPa to 1.0016 at around 2 GPa, as shown in Fig. 5(c). After 2 GPa pressure, the c/a ratio starts to increase. The tetragonality increases significantly at about 8.3 GPa. As the octahedral deformation and the tetragonality increase with pressure, we can expect that there is always a local spontaneous electric polarization in the system. In addition, the formation of ordered $TiO_6$ octahedra in the cubic phase creates carrier delocalization. Hence, we predict that the phase coexistence in EBTO leads to the ferroelectric-like semi-metallic state at high pressure.\\
\subsection{High-pressure Raman study}
For the complimentary investigation of EBTO under pressure, we turn to the results of the Raman spectroscopy measurements. As shown in Fig. 6(a), the ambient Raman scattering response of ferroelectric EBTO is characterized by five observable phonon modes with a noticeable diffusive low-frequency electronic (LFE) response. The observed LFE scattering response can be well understood within a collision-limited model \cite {marrocchelli2007pressure, congeduti2001anomalous, katsufuji1994electronic,
contreras1985raman, contreras1985raman1, jain1976electronic, sahu2022pressure}. The Raman scattering response of EBTO in the frequency range 90-1200 $cm^{-1}$ can be fitted with the spectral response \cite{congeduti2001anomalous},
\begin{equation}
	S(\omega)= [1+n(\omega)] [\frac{A\omega/\tau}{\omega^2+1/\tau^2} + \sum_{1}^{n}\frac{A_i\omega\Gamma_i}{(\omega^2-\omega_i^2)^2+\omega^2\Gamma_i^2}]
\end{equation}\\
\noindent $[1+n(\omega)]$ is the Bose-Einstein thermal factor. The first right-hand side term is meant to account for the collision-dominated low-frequency electronic response (dashed line) due to incoherent hopping of the electrons \cite{contreras1985raman, contreras1985raman1, jain1976electronic, sahu2022pressure, fluegel2015electronic, choi2003raman}.
$1/\tau$ is the carrier scattering rate. The sum, in the second term, represents the five phonon modes (dashed-dot line) at ambient pressure associated with the peak frequency $\omega_i$, amplitude $A_i$ and line width $\Gamma_i$. Our data fit extremely well to the above equations as shown in Fig. 6(a).\\ 
First, we look at the phonon band response to pressure. At ambient, EBTO is in a ferroelectric phase with a tetragonal P4mm symmetry. Hence, $\Gamma_{C_{4v}}^{optic} = 3A_1+4E+3B_1$ for the said crystal symmetry \cite{venkateswaran1998high}. All $A_1$ and $E$ modes are Raman as well as infrared active. $B_1$ mode is only Raman active. The $A_1$ and $E$ modes further split into the longitudinal and transverse optical (LO and TO) modes due to the long range electrostatic force in the crystal. The observed Raman peaks are well matched to the literature \cite {kumari2021systematic,pokorny2011use,sun2017crystalline,devi2015spectroscopic}. A broad peak is seen at 251 $cm^{-1}$ [$A_1(TO)$]. A very sharp peak is observed at 303 $cm^{-1}$ [$B_1, E(TO+LO)$], considered as a signature of the tetragonal phase. An asymmetric peak near 515 $cm^{-1}$ [$A_1$, E(TO)] and a broad peak centered at 720 $cm^{-1}$ $[E(LO)$, $A_1(LO)$] are also observed at ambient Raman spectrum as shown in Fig. 6(a). We have indexed all the modes as described by Venkateswaran et al.\cite {venkateswaran1998high}. A sharp peak at 837 $cm^{-1}$ appears in Eu doped $BaTiO_3$, which is not observed in pure $BaTiO_3$. This mode is also observed in $BaTiO_3$ doped with other $RE^{3+}$ ions like $Nd^{3+}$ or $La^{3+}$ \cite {pokorny2011use,sun2017crystalline,devi2015spectroscopic}. For simple perovskites, there is an octahedral breathing mode, $A_{1g}$ around 800 $cm^{-1}$, which is Raman inactive because of its symmetrical nature \cite{pokorny2011use}. But for EBTO, the $A_{1g}$ mode becomes Raman active due to the deformation of $TiO_6$ octahedron. The charge difference of two dissimilar ions at the equivalent site of EBTO due to the incorporation of $Eu^{3+}$ ion results in the deformation of $TiO_6$ octahedra.\\

Now, we will discuss the general observation of the pressure evolution of the phonon modes, as shown in Fig. 6(b). The intensity of all the low-frequency modes below 600 $cm^{-1}$ reduces with pressure and almost disappears at 9.5 GPa. But there is a significant increase in intensity with the pressure of the broad peak at 720 $cm^{-1}$ and the octahedral breathing mode centered at 837 $cm^{-1}$. The intensity of these modes reaches to a maximum value at around 12 GPa. The frequency positions of all the significant Raman modes are plotted as a function of pressure in Fig. 7(a). Peak splitting or disappearance, as well as slope discontinuities, are seen at about 2, 5.1 and 9.8 GPa (Fig. 6(b) and Fig. 7(a)), indicating structural rearrangement in polycrystalline EBTO \cite{venkateswaran1998high}. For visual clarity, the frequency positions of phonon modes centered at 251 $cm^{-1}$, 515 $cm^{-1}$ and 720 $cm^{-1}$ are individually plotted in Fig. 7(b), (c) and (d) as a function of pressure. Under increasing pressure, the broad phonon mode $A_1(TO)$ at 251 $cm^{-1}$ shows a red shift up to 1.8 GPa, beyond which it shows a blue shift with a significant slope change around 4.1 GPa as shown in Fig. 7(b). After 5 GPa, it becomes broader and could not be fitted by a single peak. We get a new broad peak at 304.5 $cm^{-1}$ after 5.1 GPa, which shifts to the higher frequency with pressure, as shown in Fig. 7(a). The sharp peak centered at 303 $cm^{-1}$, representative of ferroelectric phase \cite{poojitha2019effect}, is almost not shifted with pressure but the intensity gradually decreases and disappears above 1.8 GPa. This indicates that ferroelectricity, due to tetragonality, is not stable under pressure. The asymmetric $A_1$, $E(TO)$ peak ($\sim$ 515 $cm^{-1}$) softens up to 5.1 GPa with a slope change of around 2.2 GPa. After 5.1 GPa, there is an abrupt hardening with a total frequency shift of $15 cm^{-1}$, as shown in Fig. 7(c). In Fig. 7(a), a second peak is seen above 1.8 GPa on the high-frequency asymmetric tail of the said asymmetric mode. These two modes became more distinct upon pressure and around 4.5 GPa; they look like two different modes. After 5.1 GPa, we see another broad peak. All three modes disappear above 9.5 GPa. In Fig. 7(d), the $[E(LO)$, $A_1(LO)$] phonon frequency at 720 $cm^{-1}$, associated with the out-of-plane oxygen vibrations \cite{an2015assignment} shows a linear pressure dependence, again with significant slope changes at 1.8 GPa and 4.5 GPa.  This mode's intensity increases and becomes maximum at around 12 GPa, as shown in Fig. 6(b). With further increasing pressure, the peak becomes broad and weak. The octahedral breathing mode centered at 837 $cm^{-1}$ does not show any significant pressure-induced frequency shift, indicated in Fig. 7(a). As seen in Fig. 6(b), the breathing mode becomes more evident as pressure increases, indicating an increase in internal deformation of $TiO_6$ octahedron \cite{lu2011self,kchikech1994electronic}. In Fig. 8(a), it is worth noting that the $[E(LO)$, $A_1(LO)$] phonon peak broadens steadily up to 1.8 GPa, after which there is a strong decline of full width at half maximum (FWHM) up to 9.3 GPa. Beyond 9.3 GPa, the mode gradually broadens. Fig. 8(b) shows the line width of the breathing mode, $A_{1g}$ remains almost constant up to 2.3 GPa and jumps certainly to the maxima at 5.1 GPa, followed by an abrupt narrowing up to 10 GPa. After 10 GPa, it becomes progressively broader with pressure. A decrease in phonon mode bandwidth with pressure is associated with an increase in phonon lifetime due to the decrease in anharmonic scattering \cite{sahu2022pressure}. The increase in phonon lifetime indicates the presence of tetragonality in the sample at high pressure.\\
Thus, based on the pressure evolution of phonon modes of polycrystalline EBTO, we deduce three structural rearrangements in polycrystalline EBTO. The first is between 1.8 and 2.2 GPa, corresponding to the well-studied ferroelectric to a paraelectric phase transition; the second is between 4.5 and 5.2 GPa, and the last is around 10 GPa. The softening of the broad phonon mode $A_1(TO)$ at 215 $cm^{-1}$, as well as the progressive weakening and the disappearance of the sharp peak centered at 303 $cm^{-1}$, is compatible with the ferroelectric to paraelectric phase transition in $RE^{3+}$ doped $BaTiO_3$ as reported in the previous study \cite{venkateswaran1998high}. But the pressure-induced significant increase in the intensity of the high energy phonon modes centered at 720 $cm^-1$ and 837 $cm^{-1}$ reveals the gradual increase of the local structural inhomogeneity in the crystal, which indicates the sample is not completely transformed to pure cubic phase at the critical pressure of transition from parent tetragonal phase. Raman spectroscopy does not enable direct structural determination, but it is sensitive to instantaneous changes in atomic positions. The local structural deformations of $TiO_6$ octahedra caused by the charge difference between two dissimilar ions at the equivalent site become increasingly pronounced as pressure increases. The increase in local octahedral distortion is responsible for the increase in carrier localization, resulting in a local spontaneous ferroelectric polarization in EBTO at high pressure.\\  
Now we concentrate on the high pressure effect of the LFE Raman scattering response. It is observed that the LFE response turns into a flat continuum with pressure and almost disappears above 2.4 GPa, as illustrated in Fig. 9(a). In the reported literature related to the Raman studies, the metal-insulator (MI) transition of perovskite systems are characterized by a distinctive change from a collision-limited electronic diffusive scattering response to a flat continuum  \cite{yoon1998raman, liu1998probing, sahu2022pressure}. Similar features are also observed in other hole/electron doped strongly correlated oxide systems \cite{liu1998probing,yoon1998raman,gupta1996electronic,marrocchelli2007pressure,congeduti2001anomalous,katsufuji1994electronic}. This suggests, a pressure induced metallic-like state can be achieved in EBTO. Therefore, we look into the behaviour of the electronic Raman-scattering response by analyzing the diffusive low-frequency scattering response with pressure. Generally, from the crystallographic analysis, the strong lattice disorder in this system is responsible for the carrier scattering mechanism \cite{yoon1998raman, liu1998probing, billinge1996direct,radaelli1996charge,louca1997evidence}. Like other similar perovskite systems, the distortion in EBTO is primarily associated with the $TiO_6$ octahedra \cite{yoon1998raman}. The strong lattice disorder due to the octahedral distortion at ambient conditions is responsible for the large value (210 $cm^{-1}$ ) of the carrier scattering rate, as shown in Fig. 9(b). The diffusive scattering response of EBTO is quite similar to that observed in $Fe_4Nb_2O_9$ and $A_{1-x}A'_xMnO_3 (A= Pr, La$ and  $A'= Pb, Ca, Sr)$\cite{yoon1998raman, sahu2022pressure}. As a function of increasing pressure in the ferroelectric phase, the carrier scattering rate ( $1/\tau$) decreases approximately linearly with pressure. A discontinuity in the decay of carrier scattering rate is observed near 4.9 GPa.  The reducing carrier scattering rate likely reflects the reduction of long-range lattice disorder with pressure. The reduction of lattice disorder is a direct consequence of carrier delocalization and is associated with the insulator to metallic-like transition \cite{yoon1998raman}. In our system, the carrier scattering rate reduces almost by 89.5 \% (210 $cm^{-1}$ to 22 $cm^{-1}$) compared to its initial value. This suggests a pressure-induced reduction of lattice disorder, hence the charge delocalization in the system. Thus, the metallic-like state can be achieved in EBTO under high pressure. After 10.8 GPa, we can not fit the spectrum using the above equation due to broadening. The Raman data shows the presence of clusters of ferroelectric polarization inside the sample, corroborating the XRPD analysis proposing mixed phase at high pressure.\\
\subsection{Pressure dependent dielectric permittivity, dielectric loss and dc resistance}
To observe the change in ferroelectric properties of EBTO, dielectric permittivity ($\epsilon$) and dielectric loss ($\delta$) measurements were performed in the pressure range 0.5-7.1 GPa. The driving frequency varies from 50 Hz to 1 MHz at each pressure value.  Fig. 10(a) shows at low pressure (0.6 GPa) and low-frequency range, the value of the relative dielectric constant ($\epsilon$) is around 450, which is comparable with the report by Samara \cite{G. A. Samara}. $\epsilon$ decreases exponentially with increasing frequency at all pressures. Fig. 10(b) shows dielectric constant increases up to about 4.2 GPa. After 4.2 GPa, the value of $\epsilon$ decreases, and above 4.5 GPa, it drops significantly with pressure. The value of $\epsilon$ is in the order of $10^3$ at 4.2 GPa in the low-frequency range. We have not seen any linear dependency of inverse dielectric constant with pressure after 4.2 GPa. Generally, the inverse dielectric constant shows a linear pressure dependence for the paraelectric phase as described by Basu et al. \cite{A. Basu}. This suggests that under high pressure, EBTO does not exhibit a pure paraelectric phase like BT. From XRPD and Raman, we have seen the presence of phase coexistence (tetragonal (P4mm) + cubic (Pm$\bar{3}$m)) above 2 GPa, indicating a ferroelectric-paraelectric heterostructure in the system. Even though 75$\%$ of the sample gets converted to the cubic phase, we have about 25 $\%$ tetragonal phase, possibly in microscopic clusters. The local $TiO_6$ octahedral deformation caused by the incorporation of $Eu^{3+}$ ions in $BaTiO_3$ increases with pressure, indicating an increase in local ferroelectric ordering in the system at high pressure. Thus, the high-pressure paraelectric states are fragile against the presence of such localized microscopic ferroelectric order in EBTO. The relative change in the dc resistance with the ambient pressure is shown in Fig. 10(c). With increasing pressure, the dc resistance decreases and shows a minimum at around 3.5 GPa, which is very close to the pressure value where we observe a complete transition to the mixed phase with the constant phase fraction [See fig. 2(c)]. The decrease in dc resistance indicates the increase in carrier density due to delocalization, supporting a strain-induced electronic structural modification in EBTO, as observed in the pressure-induced electronic Ramna-scattering response. An increase in phase fraction of the regular cubic structure may result in the modification of the fermi surface topology resulting in an electronic topological transition with a possibility of semi-metallic type behaviour. The sharp increase in dc resistance between 3.5 to 5 GPa can then be associated with the charge localization due to the increase in tetragonality and octahedral distortion of the $TiO_6$ octahedra in the tetragonal phase after the transition pressure [See fig. 5(b)]. Such a drastic change in dc resistance, and hence in the carrier density, is unusual, but it can happen when a transition from a topological phase of a material to the strain-induced correlated phase occurs \cite{ok2021correlated}. The Dirac semi-metallic state can be achieved in this type of strain-induced symmetry-modified correlated perovskite oxide systems in the extreme quantum limit \cite{ok2021correlated}. Above 5 GPa, dc resistance reduces and almost saturates to the resistance value comparable to the present tetragonal phase. It may be due to the movement of trapped charges under the application of an electric field at high pressure \cite{R. Jana}.\\

The continuous decrease in carrier scattering rate of the electronic Raman-scattering response with pressure with a sudden drop at about 4.9 GPa, indicates the carrier delocalization, possibly leading to a metallic-like state. Anderson and Blount had proposed long back about the metallic transitions with ferroelectric order \cite{anderson1965symmetry}. Sergienko et al. \cite{sergienko2004metallic} showed that pyrochlore $Cd_2Re_2O_7$ shows metallic ferroelectricity at about 200K, driven by the lattice instability. Cai et al. \cite{cai2021pressure} have reported observation of pressure-induced emergence of the polar-metallic state in defect anti-perovskites $Hg_3Te_2X_2 (X= Cl, Br)$, and relates to the softening of the polar phonon mode. Jeong et al. \cite{jeong2011structural} have carried out a neutron total diffraction study on oxygen deficient $BaTiO_{2.75}$. They have shown the coexistence of distorted Ti-O bond (ferroelectric tetragonal phase ) and undistorted Ti-O bond (metallic cubic phase), resulting in the coexistence of the metallic-type ferroelectric phase. We believe that in the case of our sample, the presence of smaller clusters of the polar tetragonal phase in the major cubic phase leads to the development of microstructures. The coexistence of both phases at high pressure may lead to a strain-induced quantum phase transition with semi-metallic type ferroelectric behaviour in EBTO.                            
\section{Conclusion}
 We have investigated the crystal structure and the dielectric properties of rare earth Eu doped $BaTiO_3$ polycrystalline powder at high pressure by means of x-ray powder diffraction, Raman spectroscopy, and dielectric constant measurements. At ambient pressure, EBTO crystallizes in the tetragonal (P4mm) symmetry. Our XRPD results show a phase coexistence (cubic (Pm$\bar{3}$m) + tetragonal (P4mm)) above 1.4 GPa. The octahedral distortion of the $TiO_6$ octahedra and the tetragonality reduces with pressure up to 2 GPa. The $TiO_6$ octahedral deformation increases above the critical pressure with a slope change at 8.2 GPa. Beyond 2 GPa, the c/a ratio also starts to rise, peaking at 8.3 GPa. Based on the pressure-dependent phonon mode observations, we deduce structural rearrangements in polycrystalline EBTO at around 2, 5, and 10 GPa. The softening, weakening, and disappearance of the low-frequency Raman modes reveal the ferroelectric tetragonal to the paraelectric cubic phase transition. But the significant increase in the intensity of the high-frequency phonon modes with pressure indicates the local structural inhomogeneity in the crystal. The increase in local structural deformation of $TiO_6$ octahedra with pressure results in the increase in the intensity of the octahedral breathing mode centered at 837 $cm^{-1}$. So high-pressure ferroelectric phase can be obtained locally in EBTO. The LFE response suggests the pressure-induced charge delocalization due to the reduction of lattice disorder, possibly leading to a metallic-like state. High-pressure dielectric constant and dc resistance data can be explained by the presence of localized microscopic ferroelectric ordering in EBTO. Dirac semi-metallic state can be achieved in EBTO due to the strain-induced electronic structural modification. Our results suggest the ferroelectric-like semi-metallic state in EBTO at high pressure.                
\begin{acknowledgments}
 The authors gratefully acknowledge the financial support from the Department of Science and Technology, Government of India, to visit XPRESS beamline in the ELETTRA Synchrotron light source under the Indo-Italian Executive Programme of Scientific and Technological Cooperation. MS gratefully acknowledges the CSIR, Government of India, for the financial support to carry out the PhD work.	 
\end{acknowledgments}
\noindent
{\bf {Author Declarations}} 
\noindent All authors have an equal contribution. All authors reviewed the manuscript.

\noindent
{\bf {Conflict of interest:}} The authors declare no competing no conflict of interest.
 
 \pagebreak
 \begin{figure}[htb]
 	\begin{center}
 		\includegraphics[height=7in,width=7in]{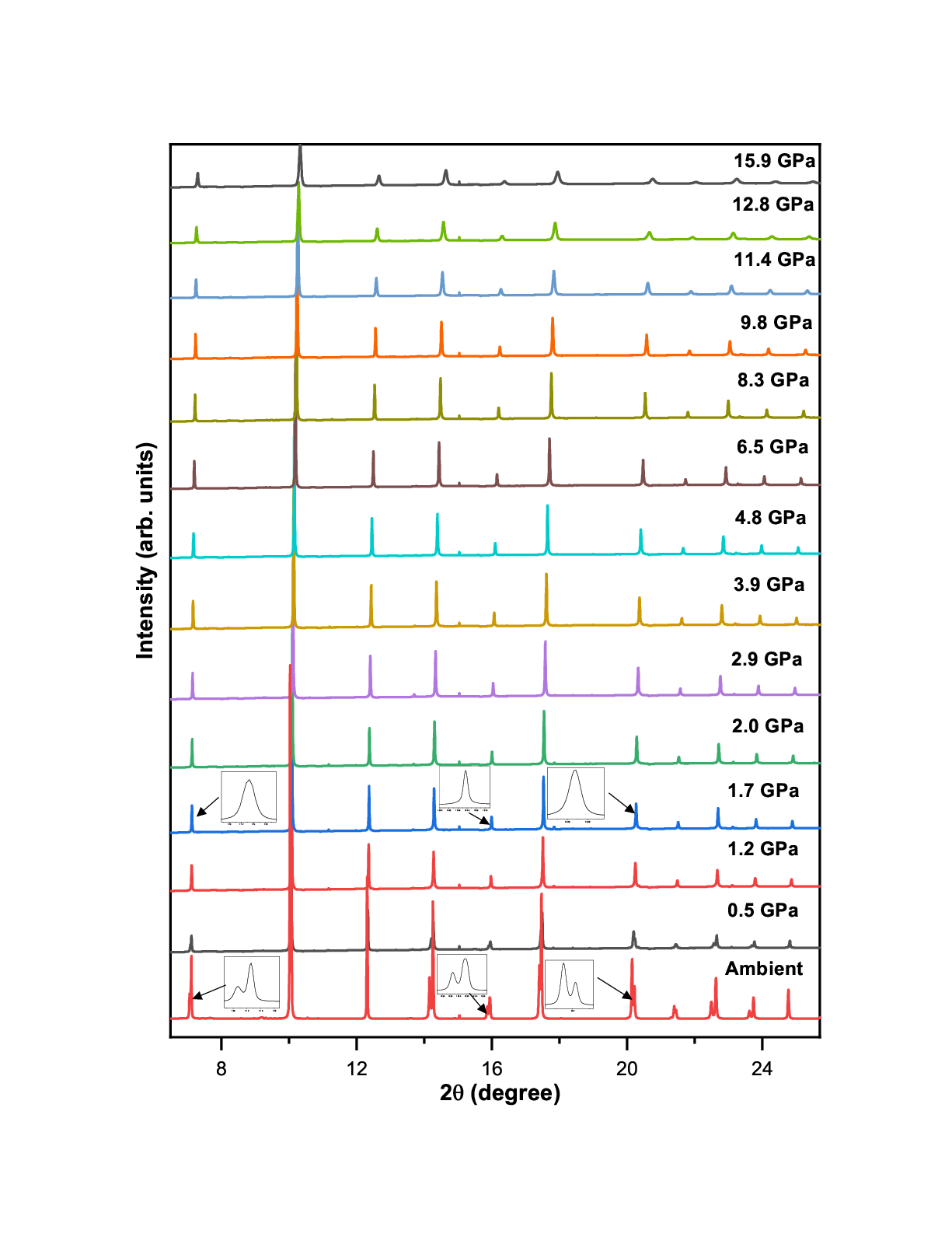}
 		\caption{Pressure dependent XRPD patterns of EBTO at selected pressures. All the Bragg profiles indicating the tetragonal phase, transform into singlets at 1.7 GPa, as indicated in the insets.}
 	\end{center}
 \end{figure}
 \begin{figure}[htb]
	\begin{center}
		\includegraphics[height=7in,width=4in]{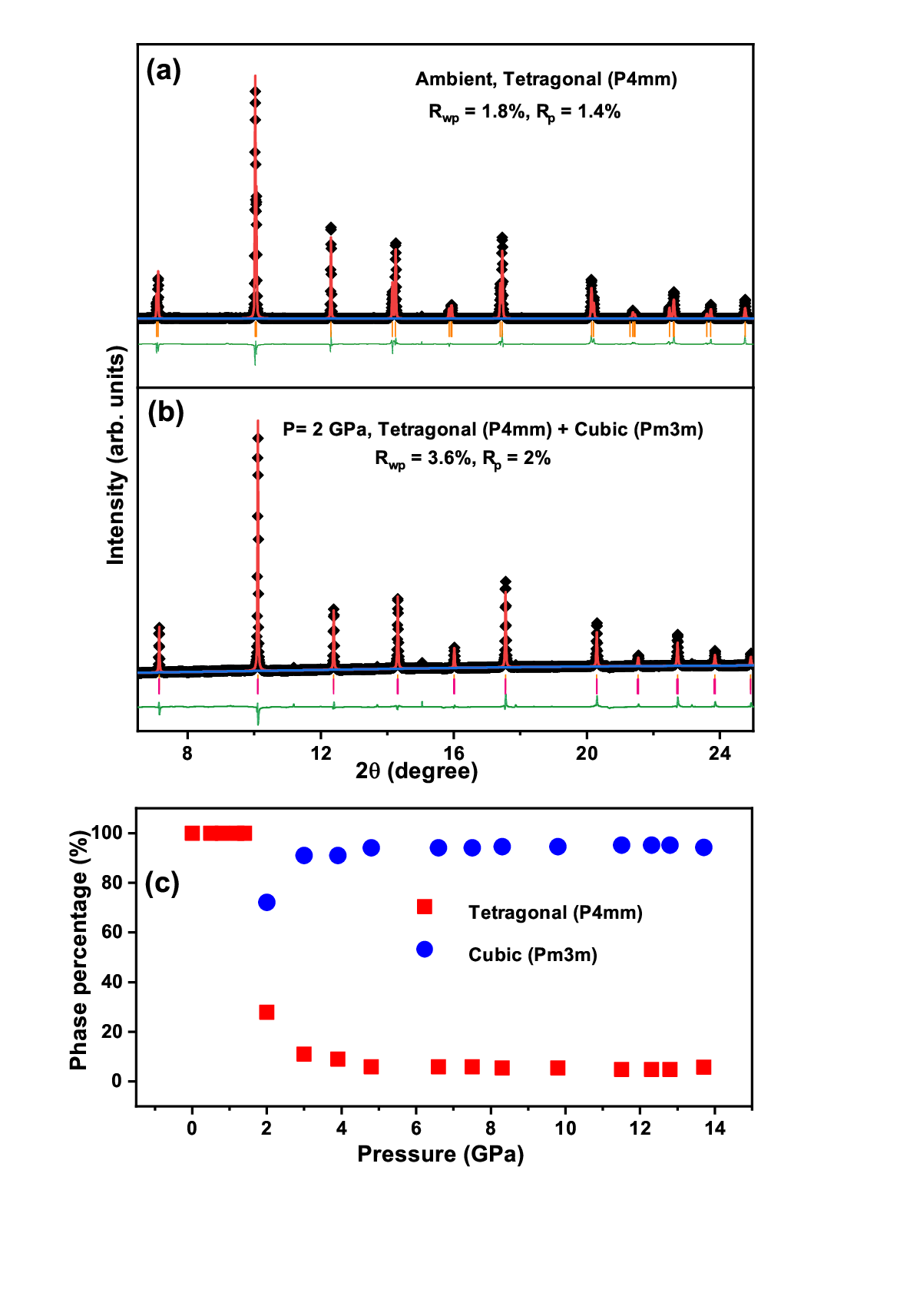}
		\caption{Rietveld refinement of x-ray diffraction patterns at (a) ambient pressure (tetragonal phase, SG: P4mm) (b) 2 GPa, fitted with the phase coexistence model (Cubic (SG: Pm$\bar{3}$m) + tetragonal (SG: P4mm) ). The bold black diamonds are the observed data points. The red line over the data point shows the fit to the data; the blue line represents the background, and the difference between observed and fitted data is shown by the green line below. The Bragg positions are marked by vertical bars. (c) Phase percentage as a function of pressure. }
	\end{center}
\end{figure}        
\begin{figure}[htb]
	\begin{center}
		\includegraphics[height=7in,width=7in]{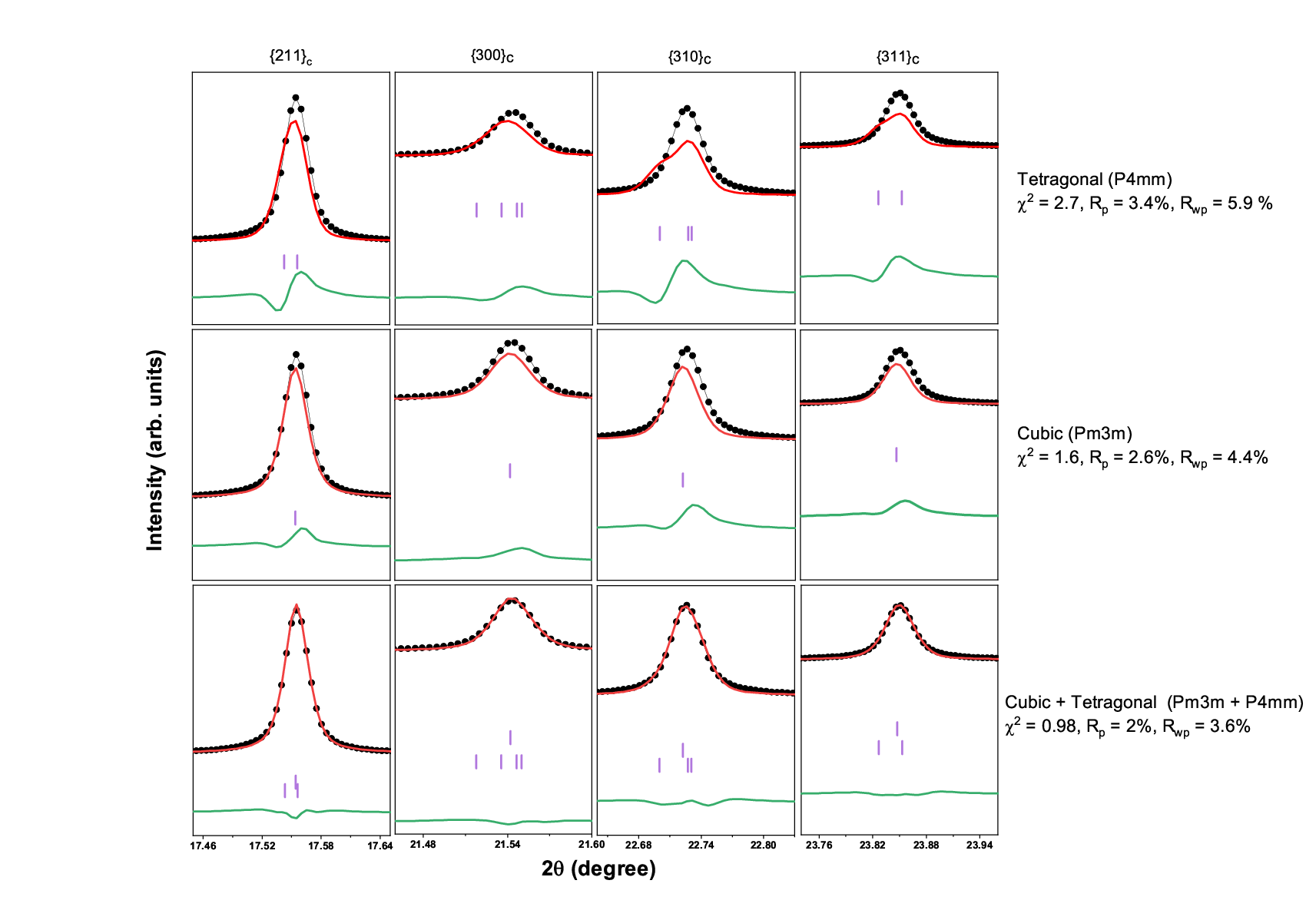}
		\caption{The Rietveld refinement of the diffraction pattern at 2 GPa with the Tetragonal (P4mm), Cubic (Pm$\bar{3}$m), Cubic (Pm$\bar{3}$m) + Tetragonal (P4mm) structural models in a restricted $2\theta$ range for clear visibility. The black dots are the observed data points. The red line over the data point represents the fit to the data, and the continuous line at the bottom depicts the difference between observed and calculated lines. In the Cubic (Pm$\bar{3}$m) + Tetragonal (P4mm) structural model fit, the upper and lower vertical lines correspond to the Bragg lines of the Cubic (Pm$\bar{3}$m) and Tetragonal (P4mm) phase, respectively}
	\end{center}
\end{figure}
\begin{figure}[htb]
	\begin{center}
		\includegraphics[height=7in,width=7in]{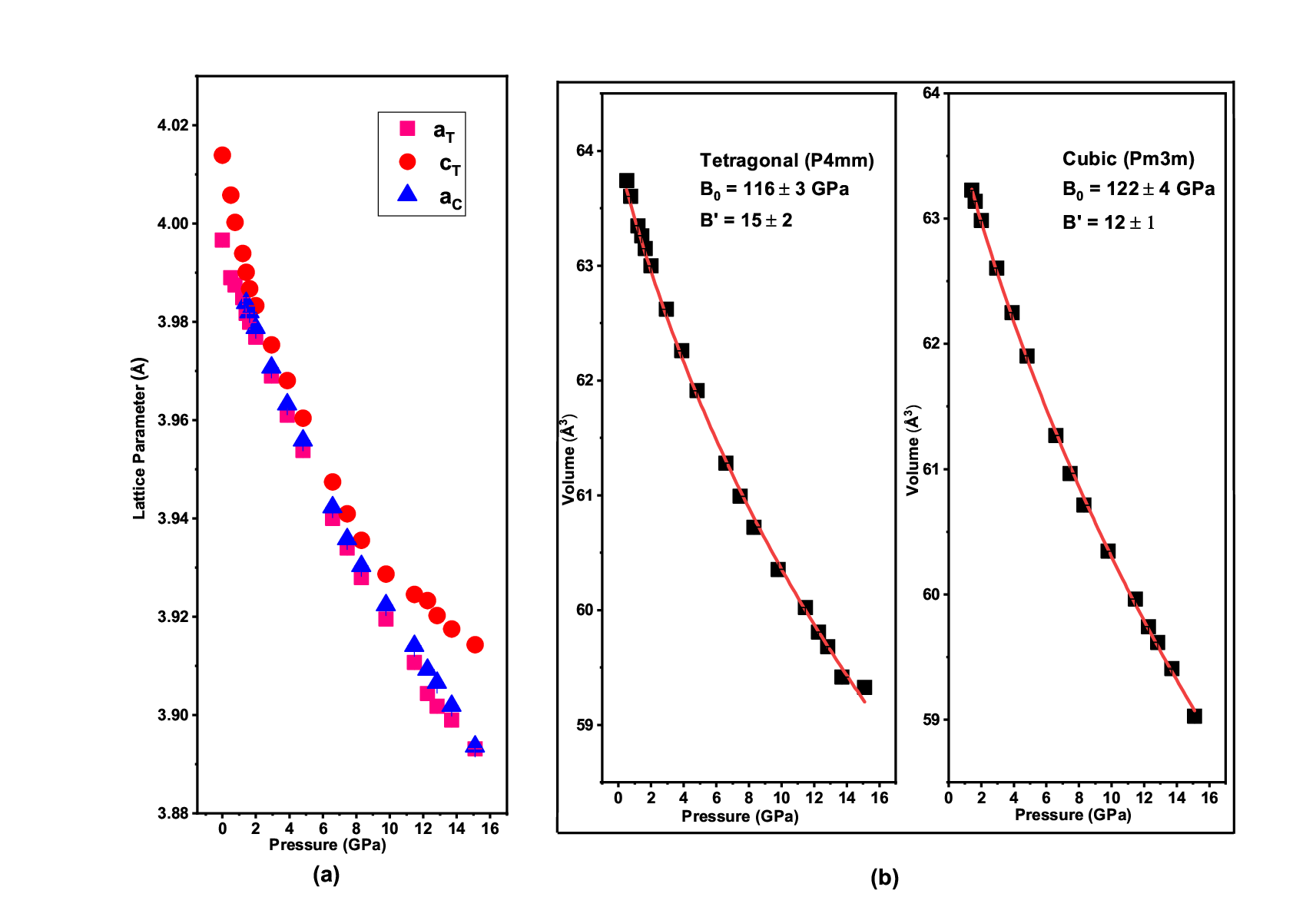}
		\caption{[a] Evolution of lattice parameters of EBTO with pressure. The pink square and the red circles denote the `a' and `c' lattice parameters respectively of the tetragonal (P4mm) phase, while the blue triangles represent the lattice parameter a of the cubic (Pm$\bar{3}$m) cell. [b] Pressure evolution of unit cell volume of polycrystalline EBTO. Solid lines passing through the data are Birch-Murnaghan equation of state fit to the volume data.}
	\end{center}
\end{figure}   
 \begin{figure}[htb]
 	\begin{center}
 		\includegraphics[height=7in,width=7in]{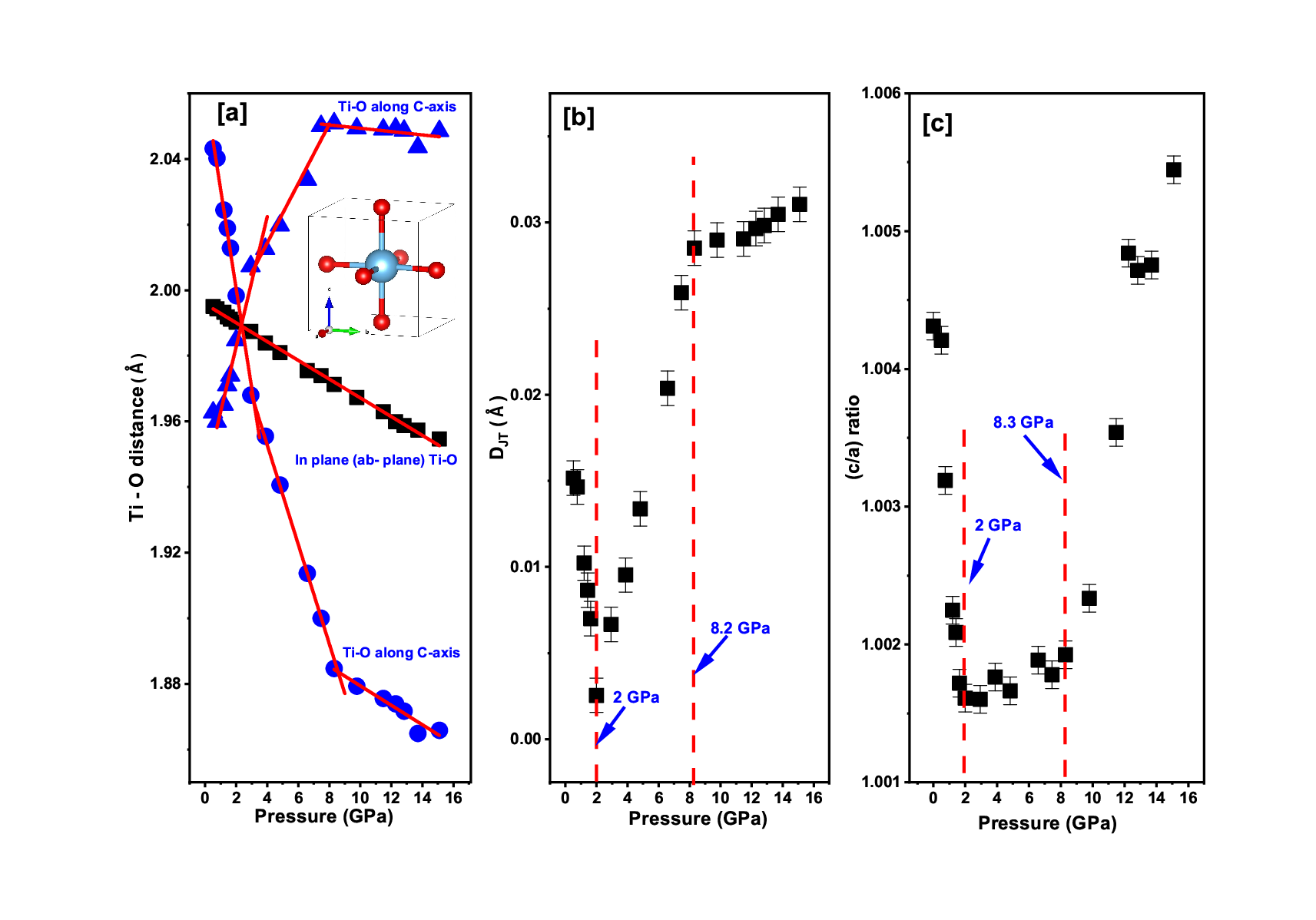}
 		\caption{[a] Pressure dependence of all six Ti-O bond distances in $Ti-O_6$ octahedra at ambient temperature. Four in-pane (ab-plane) Ti-O bond distances, one long Ti-O bond distance and one short Ti-O bond distance along `c'-axis are represented by black squares, blue circles and triangles, respectively. The inset represents the $Ti-O_6$ octahedra, where the grey and red spheres represent the $Ti^{4+}$ ions and $O^{2-}$ ions, respectively. [b] Pressure evolution of Jahn-Teller distortion of the $Ti-O_6$ octahedra. Anomaly observed at 2 GPa and 8.2 GPa. [c] Pressure dependence of the c/a ratio (tetragonality). Certain anomalies are observed at 2GPa and 8.3 GPa. }
 	\end{center}
 \end{figure} 
\begin{figure}[htb]
	\begin{center}
		\includegraphics[height=7in,width=5in]{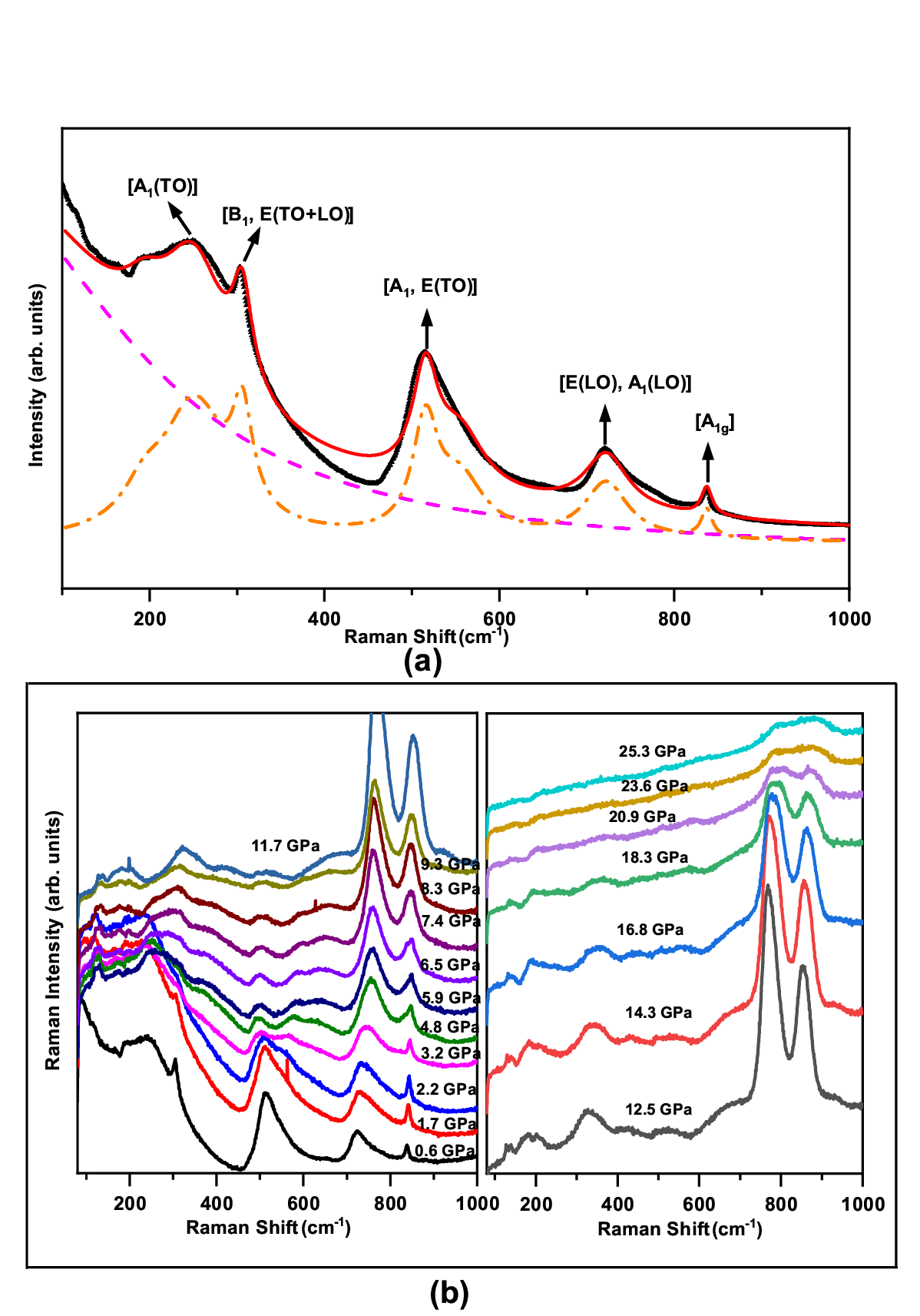}
		\caption{(a) The deconvolution of ambient Raman spectrum along with the fit using the collision-limited model [Eq. 3]. The solid red line is the best fit to the data. The contribution of Phonon modes (dot-dashed line) and LFE contributions (dashed line) are shown separately. (b) Evolution of Raman spectra at selected pressure points.}
	\end{center}
\end{figure} 
 \begin{figure}[htb]
 	\begin{center}
 		\includegraphics[height=7in,width=7in]{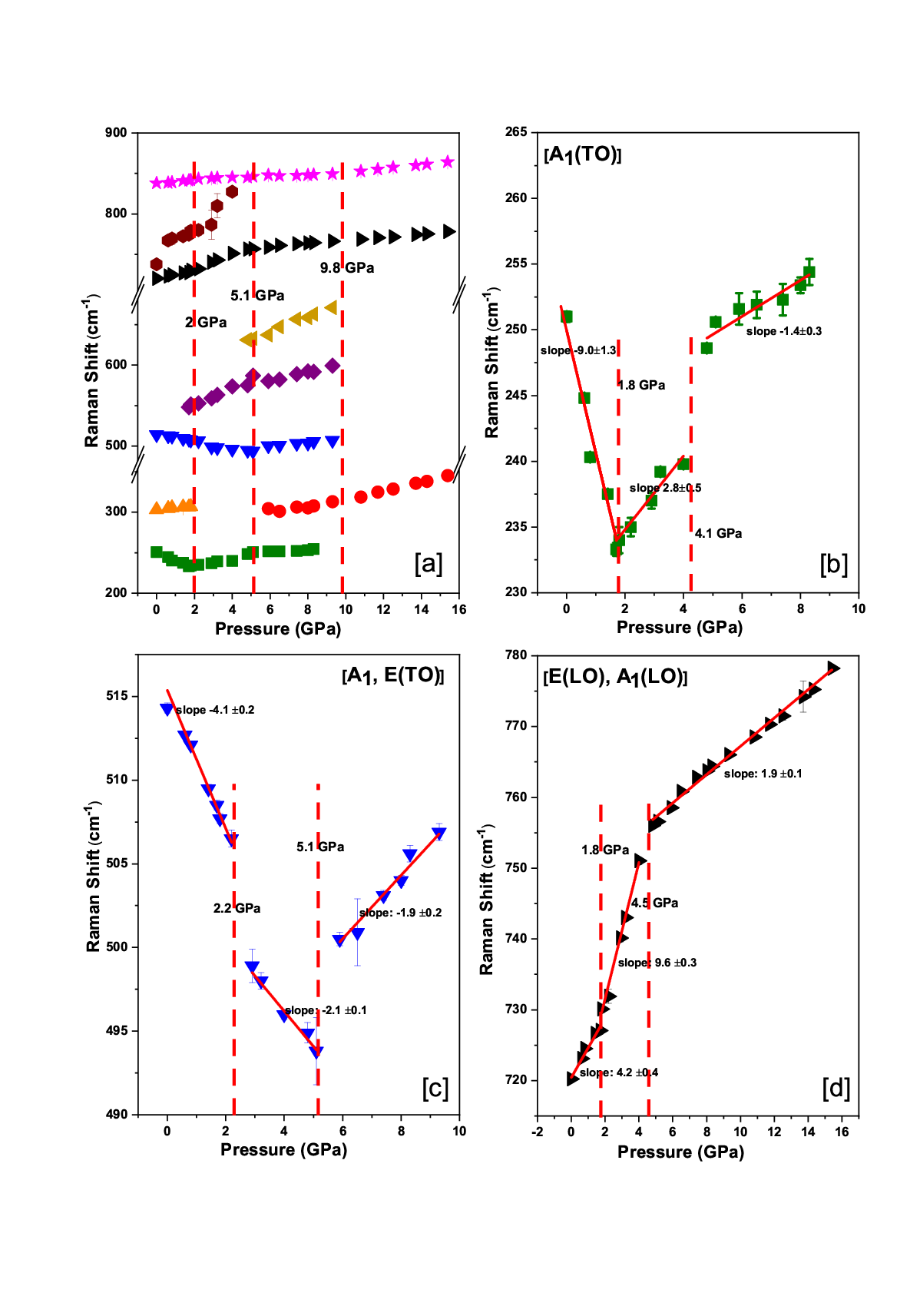}
 		\caption{[a] Raman mode values observed in polycrystalline EBTO plotted as a function of pressure. Slope discontinuities, splitting or disappearance of phonon modes are observed at 2 GPa, 5.1 GPa and 9.8 GPa, indicating structural rearrangement in the crystal. For visual clarity, the frequency of the phonon modes [b] $[A_1(TO)]$, [c] $[A_1, E(TO)]$ and [d] $[E(TO), A_1(LO)]$ are plotted as a function of pressure. The modes $[A_1(TO)]$ and $[A_1, E(TO)]$ show red shift up to 1.8 GPa and 5.1 GPa, respectively. All the modes show significant slope changes around the said pressure values.}
 	\end{center}
 \end{figure}
\begin{figure}[htb]
	\begin{center}
		\includegraphics[height=7in,width=5in]{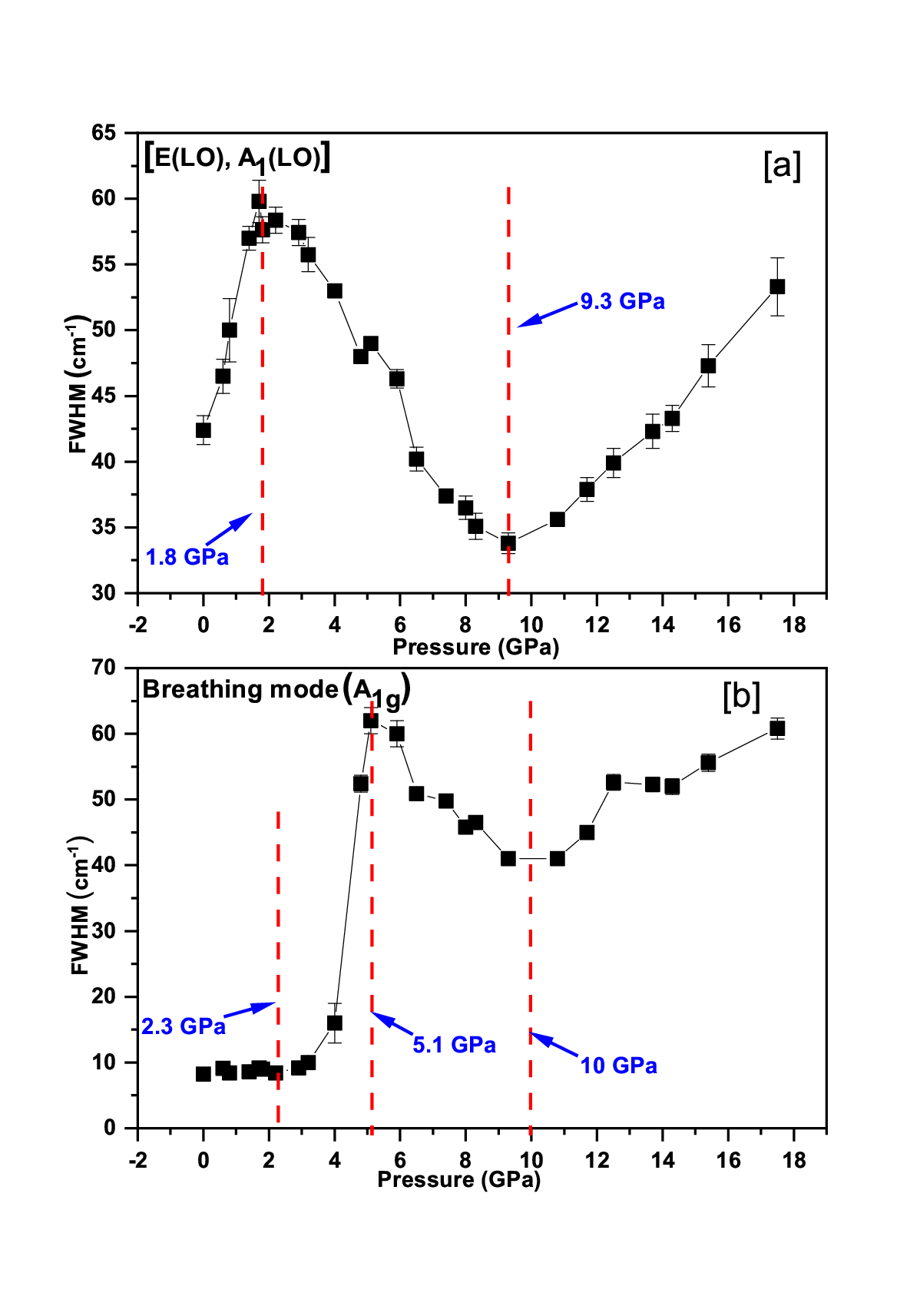}
		\caption{Pressure dependent FWHM of (a) $[E(TO), A_1(LO)]$ phonon mode. Anomalies were observed at 1.8 GPa and 9.3 GPa, as shown in the figure. (b) Breathing mode, $A_{1g}$. As illustrated in the figure, anomalies were detected at 2.3 GPa, 5.1 GPa and 10 GPa.}
	\end{center}
\end{figure}
\begin{figure}[htb]
	\begin{center}
		\includegraphics[height=7in,width=7in]{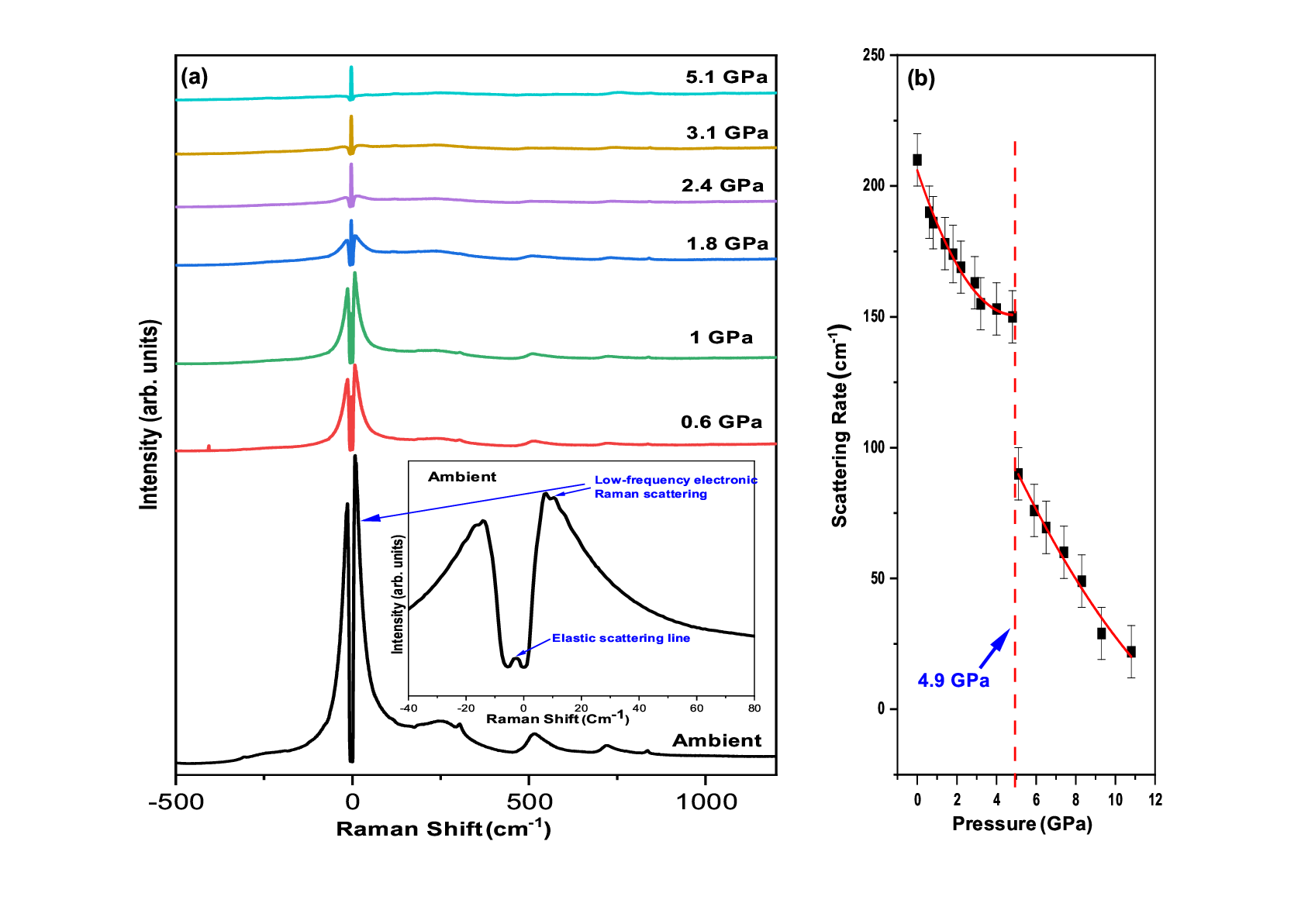}
		\caption{(a) Raman spectra of polycrystalline EBTO for several pressures from ambient to 5.1 GPa. We have taken the data using the Bragg filter to see the pressure effect of the low-frequency response near the elastic line. Low-frequency electronic Raman scattering response and the elastic line are shown in the inset for visual clarity. LFE response turns into a flat continuum beyond 2.4 GPa. (b) Diffusive low-frequency electronic scattering rate to the Raman spectra at different pressures. There is a discontinuity of the decay of carrier scattering rate at 4.9 Gpa.}
	\end{center}
\end{figure}
\begin{figure}[htb]
	\begin{center}
		\includegraphics[height=7in,width=6in]{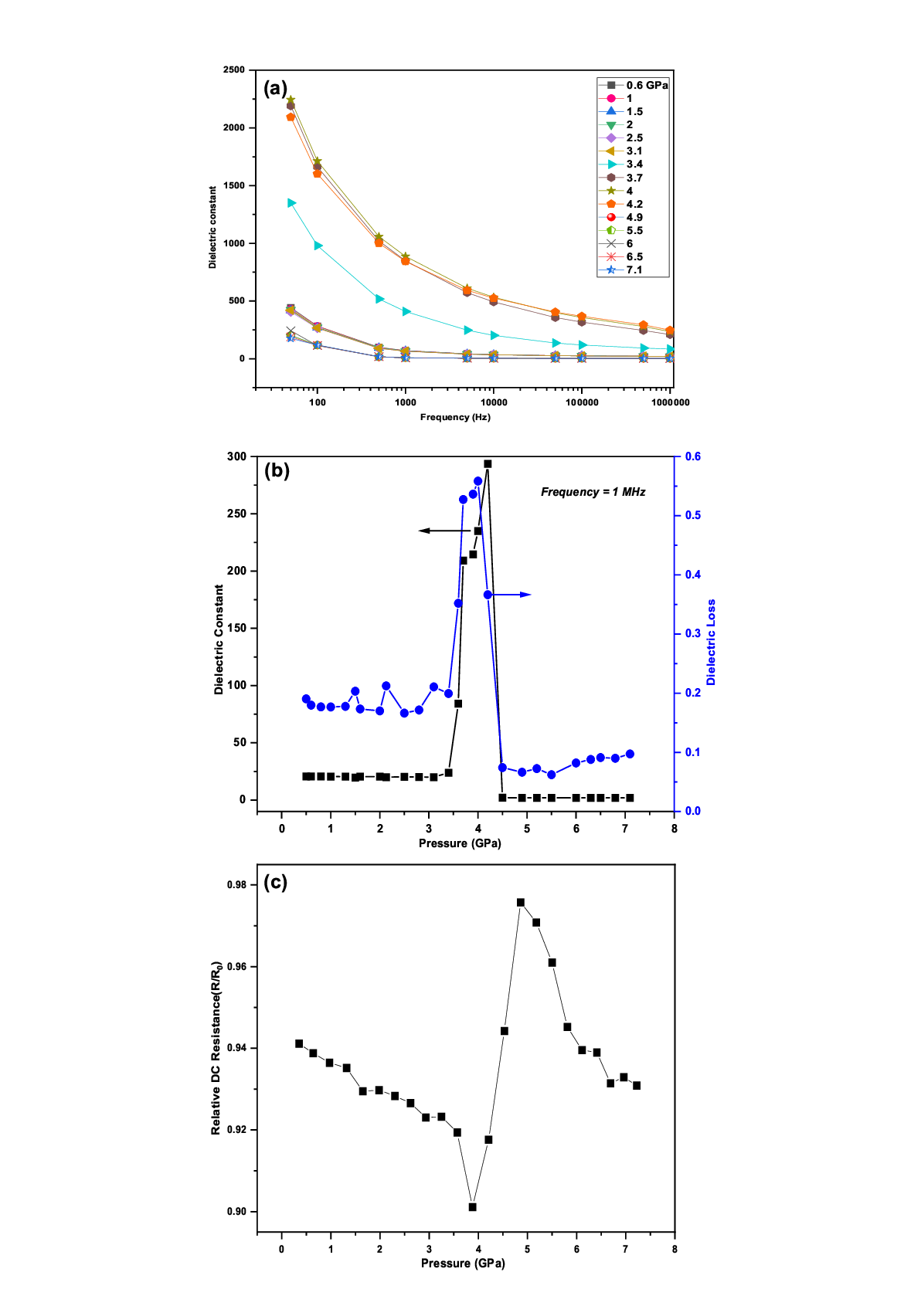}
		\caption{(a) Frequency dependent dielectric constant ($\epsilon$) of EBTO at selected pressures. (b) Pressure evolution of dielectric constant (squares) and dielectric loss (dots) of EBTO at 1 MHz. (c) Relative change in dc resistance with pressure.}
	\end{center}
\end{figure}
\end{document}